# A Hail Size Distribution Impact Transducer

John E. Lane[a], Robert C. Youngquist[b], William D. Haskell[a], and Robert B. Cox[a]

[a]*ASRC Aerospace Corporation, P.O. Box 21087, Kennedy Space Center, FL 32815*
*John.Lane-1@ksc.nasa.gov    William.Haskell-1@ksc.nasa.gov    Robert.Cox-1@ksc.nasa.gov*

[b]*National Aeronautics and Space Administration (NASA), Kennedy Space Center, FL 32899*
*Robert.C.Youngquist@nasa.gov*

**Abstract:** An active impact transducer has been designed and tested for the purpose of monitoring hail fall in the vicinity of the Space Shuttle launch pads. An important outcome of this design is the opportunity to utilize frequency analysis to discriminate between the audio signal generated from raindrop impacts and that of hailstone impacts. The sound of hail impacting a metal plate is subtly but distinctly different than the sound of rain impacts. This useful characteristic permits application of signal processing algorithms that are inherently more robust than techniques relying on amplitude processing alone in the implementation of a hail disdrometer.

## 1. Introduction

Impact disdrometers have long been a useful meteorological tool to measure and quantify rainfall drop size distributions, where a drop momentum is converted into a single electrical impulse.[1,2] The electrical impulse amplitude is converted to an estimate of drop diameter by means of an empirical calibration formula. The calibration may be a one-time procedure involving dropping numerous known-size-calibration drops from a tower of sufficient height to achieve terminal velocity. The test drop sizes must adequately span the size range of interest. Alternatively, or as a supplement to the single drop calibration, an *in situ* method of comparing the sum of accumulated drop impulses to the tip time interval of a tipping-bucket rain gauge may be utilized.[3]

Prior to launch, the Space Shuttle and other NASA launch vehicles are primarily large thermos bottles containing substantial quantities of cryogenic fuels. Because thermal insulation is a critical design requirement, the external wall of the launch vehicle fuel tank is covered with an insulating foam layer. This foam is fragile and can be damaged by very minor impacts, such as that from small- to medium-size hail, which may go unnoticed. In May 1999, hail damage to the top of the External Tank (ET) of STS-96 required a rollback from the launch pad to the Vehicle Assembly Building (VAB) for repair of the insulating foam. Because of the potential for hail damage to the ET while exposed to the weather, a vigilant hail sentry system using impact transducers can be deployed around the launch pads as a permanent warning system to record and quantify hail events, and to initiate a thorough inspection of the ET.

Since hail distributions are far less dense than raindrop distributions,[4] the surface area of a hail sensor must be much larger than that of a rainfall sensor. Overly large surface area is a detriment since the resulting overlap in drop impulses would make the instrument useless during intense storms. If the sensor collection area is too small, the measurements do not represent a good statistical sampling of the hydrometeor size distributions. Since the electrical impulse amplitude is well correlated to hydrometeor size, it is straightforward to measure and estimate hail size. However, small hailstones may be smaller than large raindrops. Therefore, impulse amplitude alone is insufficient for discriminating hail from rain in that size overlap region, a hail size regime common to the Florida climate.

A simple yet effective passive hail transducer is made using a 1-ft$^2$, 1-in-thick Styrofoam panel, covered with thick aluminum foil.[5,6] This design is used extensively by the Colorado State University Department of Atmospheric Sciences and by over 3,000 volunteers of the Community Collaborative Rain and Hail Study (CoCoRaHS) to measure and monitor hail in Colorado and adjacent states.[7] The Kennedy Space Center (KSC) hail transducer shown in Figure 1 is fabricated of sheet aluminum.[8] The shallow pyramid design persuades hail to bounce away from the sensor so that multiple hits from a single hailstone are not erroneously recorded. The bottom side contains a piezoelectric ceramic, mounted on a plate with a waterproof cover. A 1-MΩ resistor is placed in parallel with the ceramic disc to bleed off excess charge buildup when not connected to a circuit load. This design was chosen for its low cost, simplicity, and extreme durability in the launch pad environment. However, the greatest challenge in measuring hail is discriminating it from rain.

**2. Math model**

The NASA/ KSC hail impact transducer is essentially a stiff, sheet aluminum drumhead, where the primary resonance modes are centered on a 4-kHz band. There are numerous modes of vibration, closely spaced in frequency and broad enough in bandwidth to be considered a continuum. However, for the purpose of this analysis, the response of the transducer will be modeled as a set of $N$ uncoupled parallel second-order systems (mass-damper-spring), each characterized by a resonant frequency $\omega_0 = \sqrt{\kappa/M}$, damping factor $d = 2\gamma M$, and modal mass $M$. The equation of motion of a single mode is therefore

$$\ddot{x} + 2\gamma \dot{x} + \omega_0^2 x = F(t)/M \qquad (1)$$

where $F(t)$ is the force of impact from a hydrometeor, and $\gamma$ is proportional to the damping factor, introduced to simplify the solution to Eq. (1). Since the complete transducer system consists of $N$ uncoupled second-order mechanical oscillators, as depicted in Figure 2, important transducer characteristics can be deduced by studying the characteristics of a single mass-damper-spring system, corresponding to an eigenvalue-eigenvector (or mode) of the vibrating membrane.

A key to solving Eq. (1) for the displacement $x(t)$ is to choose a reasonably simple yet useful description of the impulse force $F(t)$.[9] A reasonable estimate is

$$F(t) = \begin{cases} F_0 = mv(1+\eta)/\tau & t \leq \tau \\ 0 & t > \tau \end{cases} \qquad (2)$$

where $m$ is the hydrometeor mass, $v$ is the hydrometeor velocity (usually a constant *terminal velocity*), $\eta$ is the *coefficient of elasticity*, and $\tau$ is the time interval that the impulse force is nonzero. If hail is hard (i.e., not composed of slush), the coefficient of elasticity will be close to the maximum possible value of 1. The coefficient of elasticity can be determined by direct measurement of the velocity $v$ before impact and the reflected velocity $v'$ after impact (assuming one-dimensional movement for simplicity): $v' = -\eta v$, where the minus sign signifies that the final velocity is in the opposite direction.

Taking a very simplistic approach, the time interval $\tau$ that the impulse force acts on the transducer can be modeled as

$$\tau = \xi D/v \tag{3}$$

where $D$ is the hydrometeor diameter, $v$ is the initial velocity before impact, and $\xi$ is an empirical *coefficient of hardness,* which like $\eta$, varies between 0 and 1. One might be tempted to relate $\eta$ to $\xi$ by something like $\xi = 1 - \eta^B$, where $B$ is a fitting parameter. However, the main point here is that $\eta$ and $\xi$ behave as additive inverses, i.e., when $\eta$ is near 1, $\xi$ is near 0.

Applying the estimates from Eq. (2) and Eq. (3) to Eq. (1), the solution for displacement is

$$x(t) = \frac{F_0}{M\omega_0^2} \begin{cases} 1 - e^{-\eta}\left(\cos\omega_1 t + \frac{\gamma}{\omega_1}\sin\omega_1 t\right) & t \leq \tau \\ e^{-\gamma(t-\tau)}\left(\cos\omega_1(t-\tau) + \frac{\gamma}{\omega_1}\sin\omega_1(t-\tau)\right) - e^{-\eta}\left(\cos\omega_1 t + \frac{\gamma}{\omega_1}\sin\omega_1 t\right) & t > \tau \end{cases} \tag{4}$$

where $\omega_1 \equiv \sqrt{\omega_0^2 - \gamma^2}$.

Eq. (4) essentially shows that the system impulse response to a finite-width delta function does not decrease with decreasing pulse width as long as the period of the transducer response oscillation is much less than the width of the impulse: $2\pi/\omega_0 \ll \tau$. The response roll-off slope above a characteristic frequency becomes independent of hardness, where the offset of the slope is controlled by the impulse time interval, $\tau$, from Eq. (2) and Eq. (3), and is inversely proportional to the coefficient of hardness, $\xi$.

The next step is to run some typical values through the second-order system model of Eq. (4), using approximations for hydrometeor terminal velocity[10] and mass, based on hydrometeor diameter $D$: $v \approx 142 D^{1/2}$ and $m = \pi D^3 \rho /6$, where $\rho$ is the density of ice. Using the values from Table 1 in Eq. (4) results in the curves of Figure 3, corresponding to the transducer signal response, proportional to the displacement $x(t)$, versus mode frequency $f_0 = \omega_0 / 2\pi$, for various values of $\xi$ from $\xi \approx 1$ (raindrop) to $\xi \approx 0.1$ (hard hail). Note that the *magnitude response*, $x_\omega$, along the vertical axis in Figure 3 is defined as the maximum peak value of the magnitude of $x(t)$ in Eq. (4), during the impulse interval, as a function of mode frequency: $x_\omega \equiv \max \left\{ |x(t,\omega)|_{t=\text{pulsestart}}^{\text{pulseend}} \right\}$

The hydrometeor diameter $D$ used in Table 1 corresponds to a small *rice-size* hailstone, or equivalently, a large raindrop. The damping factor $\gamma$ is picked to match the decay rate seen in typical time domain plots of the transducer response. When $t = \gamma^{-1}$ or 6.67 ms, the pulse has decayed to $1/e$ of some initial maximum value. The value chosen for the second-order system mass $M$ has no real meaning here since it is just a scale factor in front of Eq. (4). The system gain of the processing system electronics has a similar effect in the model as the modal mass $M$. Both effects can be ignored as long as the parameters remain constant.

As discussed previously, the plot in Figure 3 shows that as mode frequency increases, the separation in response between small and large values of $\xi$ increases. This result is corroborated by the spectral plots of the video in Movie 1, based on the 1,024-point fast Fourier transform of a typical hailstone and a raindrop. In the case of hail (i.e., for $\xi < 0.3$), the transducer response in the frequency band above 10 kHz is much higher than for raindrops of similar momentum. This then is the key to rain-versus-hail discrimination when using this type of drop impact transducer. The method that was chosen to process the transducer signal is based on a second-order IIR high-pass filter with a cutoff of 16 kHz. It is important to note that this implies that a sufficient sample rate of at least 44 kHz or better must be used in a digital implementation of the filter.

## 3. Calibration strategy

A hail pad calibration method developed during the 1978 Alberta Hail Project relates the dent diameter $d$ to the hail diameter $D$ as an empirical second-order polynomial:[11]

$$D = a_0 + a_1 d + a_2 d^2 \qquad (5)$$

where $a_0 = 0.38$ [cm], $a_1 = 1.11$, and $a_2 = -0.04$ [cm$^{-1}$]. Note that Eq. (5) implies that the smallest detectable hail size is $\lim_{d \to 0} D = a_0$, or about 4 mm. Any hailstone smaller than $a_0$ will go undetected by the hail pad.

In order to verify Eq. (5), ice balls were fabricated, measured, and then dropped from a 20-m platform. The ice ball diameters, hail pad dents (Figure 4), and corresponding diameters calculated from the hail pad calibration formula of Eq. (5) are plotted in Figure 5 as a function of actual size. Since the slope is very close to unity, Eq. (5) is therefore validated as a means to calibrate the hail transducer using an *in situ,* side-by-side comparison with standard hail pads during naturally occurring hail events. It should be noted that Eq. (5) does not take into account hail hardness. However, experts at the CoCoRaHS analysis lab at CSU are able to estimate hail hardness by observing the type of dent pattern. For the purposes of this current work, hail pad dents, fabricated hail balls, and the resulting calibration all assume hard hail.

Data was collected from a July 7, 2004, thunderstorm containing hail that passed over KSC and the surrounding area. Both the hail transducer and hail pads colocated at three different sites were used to collect data. Site WEK, about 30 km south of KSC, measured the largest amount of hail. The other sites collected only trace amounts. The probability distribution functions (PDFs) of measured hail hits and corresponding discrete histograms for both the hail pads and the hail transducer are shown in Figures 6 and 7, respectively.

The relationship between the PDF $P(x)$ and histogram $h(x)$ of variable $x$ is $P(x)\Delta x = h(x)$, where $\Delta x$ is the width of the histogram bin. Each $k^{th}$ bin of $h(x)$ corresponds to $x$ in the range of $(k-1)\Delta x \leq x < k\Delta x$. Note that in the case of the hail transducer, $x$ is a dimensionless impulse peak amplitude, similar to $x_\omega$, but not restricted to a narrow-band frequency, scaled to have a value between 0 and 1 ($x$ should not be confused with the mechanical displacement $x(t)$ from previous discussions). This scaling is convenient for implementation on a fixed-point fractional digital signal processor where numbers are restricted to the range of -1 to 1.

The value of $x$ is dependent on many steps in the signal processing stream and can be related to a physical signal voltage through the conversion of mechanical displacement via the piezoelectric element of the transducer. Because absolute voltage units are not needed when following this type of calibration procedure, it is practical to calibrate the hail monitor system with the assumption that the steps in the signal processing remain constant and have a predictable effect on the impulse amplitude measurement unless an intentional adjustment is made in the signal processing flow.

The PDF of drop diameters from the hail pad can be used to generate a calibration curve for the hail transducer using a probability matching method.[12] Note that this procedure assumes that the number of impulses counted by the hail transducer is greater than the number counted by the hail pad. If the opposite is the case, a slightly modified procedure is followed. Since the hail pad cannot detect hail sizes smaller than $D_0 = a_0 \approx 4$ mm, the initial step in matching PDFs is to solve for the impulse value $x_0$ corresponding to $D_0$:

$$\int_{x_n}^{\infty} P(x)\, dx = \int_{D_n}^{\infty} P(D)\, dD \tag{6}$$

with $n = 0$, or in the case of the discrete histograms, consisting of $M_x$ and $M_D$ bins:

$$\sum_{k=k_n}^{M_x} h_x(k) = \sum_{j=j_n}^{M_D} h_D(j) \tag{7}$$

where $x_k \equiv k\Delta x$ and $D_j \equiv j\Delta D$. The result of the application of Eq. (7) to the July 7 WEK histograms $h(x_k) \equiv h_x(k)$ and $h(D_j) \equiv h_D(j)$ is $D_0 = 4.61$ mm and $x_0 = 0.169$. This procedure is repeated for all values of $j_n$ and $k_n$ by matching all accumulated histogram totals, bin by bin.

Figure 8 shows the result of matching all histogram bins, corresponding to this *in situ calibration* of the hail transducer. A power-law fit of the form

$$D(x) = Ax^b \tag{8}$$

where $A = 11.13$ and $b = 0.621$, provides a convenient closed-form calibration formula for all values of impulse amplitude $x$, consistent with the July 7 WEK data. Figure 9 shows the result of converting the impulse amplitude hail transducer data to equivalent hail diameters using Eq. (8). The open circles correspond to extrapolated values below the hail pad $D_0$ threshold.

## 4. Summary

The mathematical model described by Eq. (4) demonstrates that the transducer's magnitude response (maximum of electrical impulse signal) is more sensitive to the hydrometeor hardness at higher frequencies. The result is that discrimination between liquid and solid hydrometeors is best achieved in a frequency band at least two to three times higher than the main transducer resonance. A signal processing implementation to utilize this effect, when sampling at a standard frequency of 48 kHz, simply involves processing the signal with a second-order high-pass filter with a cutoff frequency of 16 kHz. The active impact transducer has the potential to measure hail stone sizes well below the lower size threshold of the passive hail pad.

Future work may include investigating the correlation between impulse frequency spectral response and hail hardness. In addition, incorporating hail hardness into the calibration scheme is an important area of future work. In the case of drop tower calibration, fabricating hail balls with varying degrees of hardness might be accomplished by using small traces of impurities in the water before freezing.[13] Undoubtedly, ignoring hail hardness could lead to large errors in hail size distribution estimates from both the passive hail pad and the active hail transducer.

Table 1. Parameters value estimates used to generate Figure 3 with Eq. (4).

| $\gamma$ [s$^{-1}$] | $\eta$ | $M$ [kg] | $D$ [m] |
|---|---|---|---|
| 150 | 0.9 | 0.01 | 0.005 |

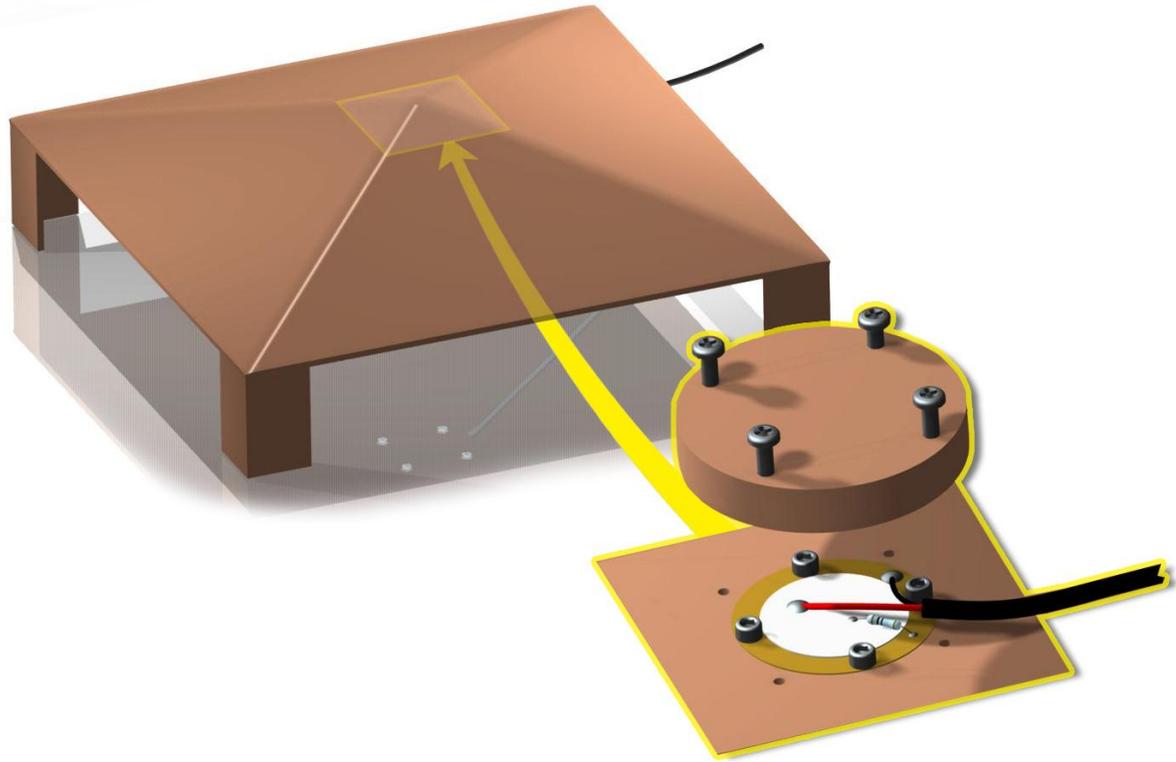

Figure1. NASA/KSC hail transducer. The piezoelectric ceramic disc is shown in white and is mounted on a brass disc substrate.

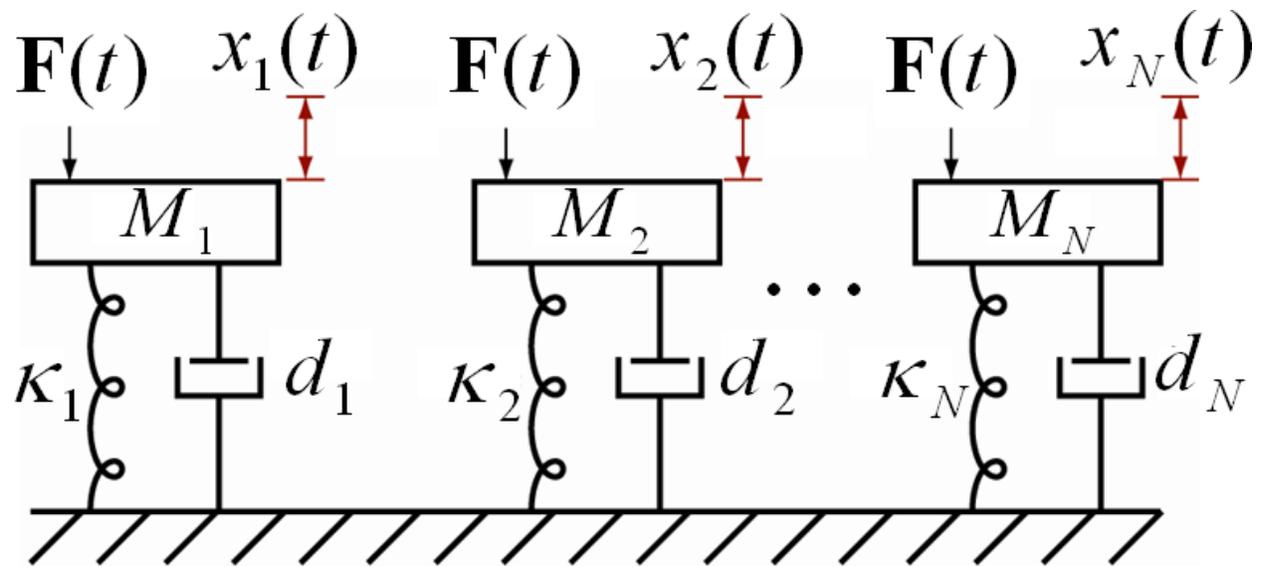

Figure 2. Lumped parameter modal transducer model.

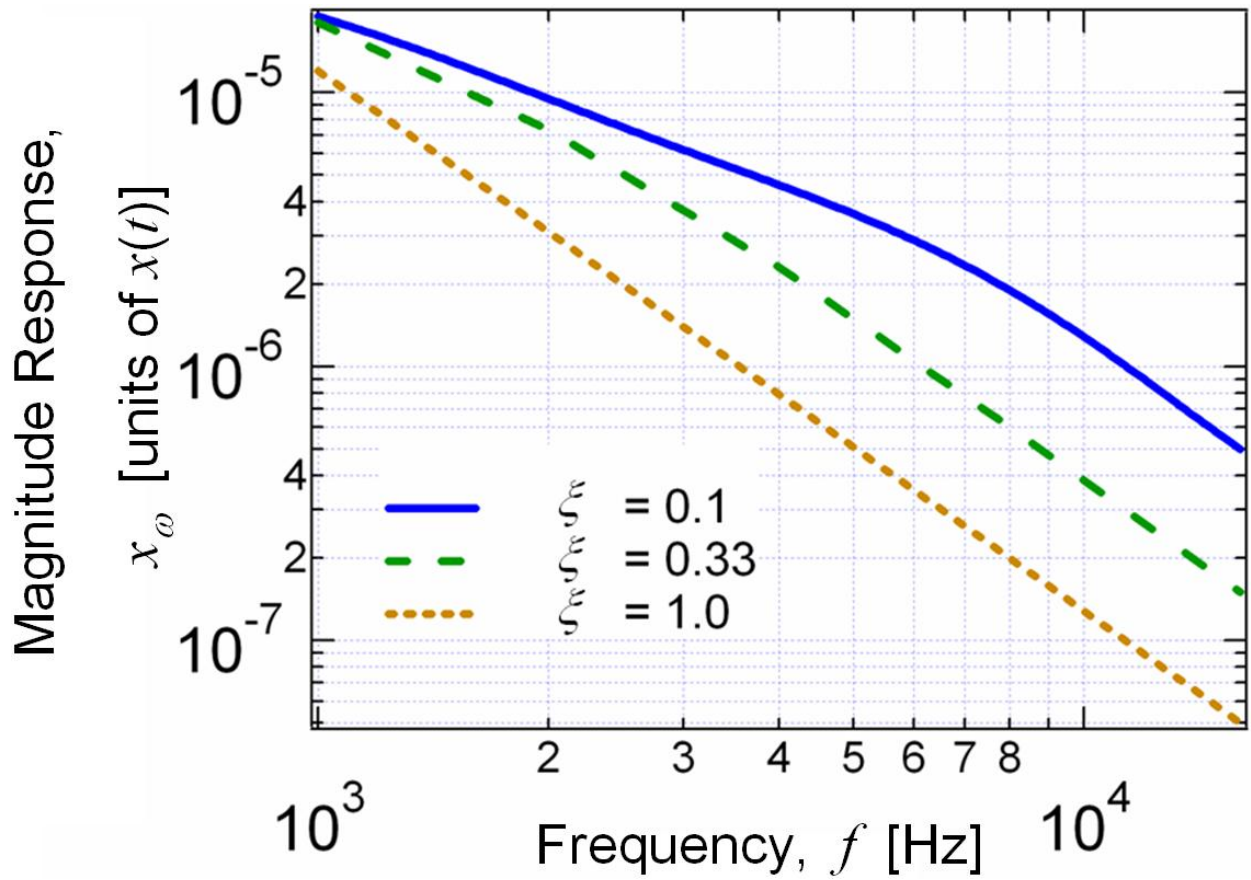

Figure 3. Model response versus frequency for various values of hydrometeor hardness.

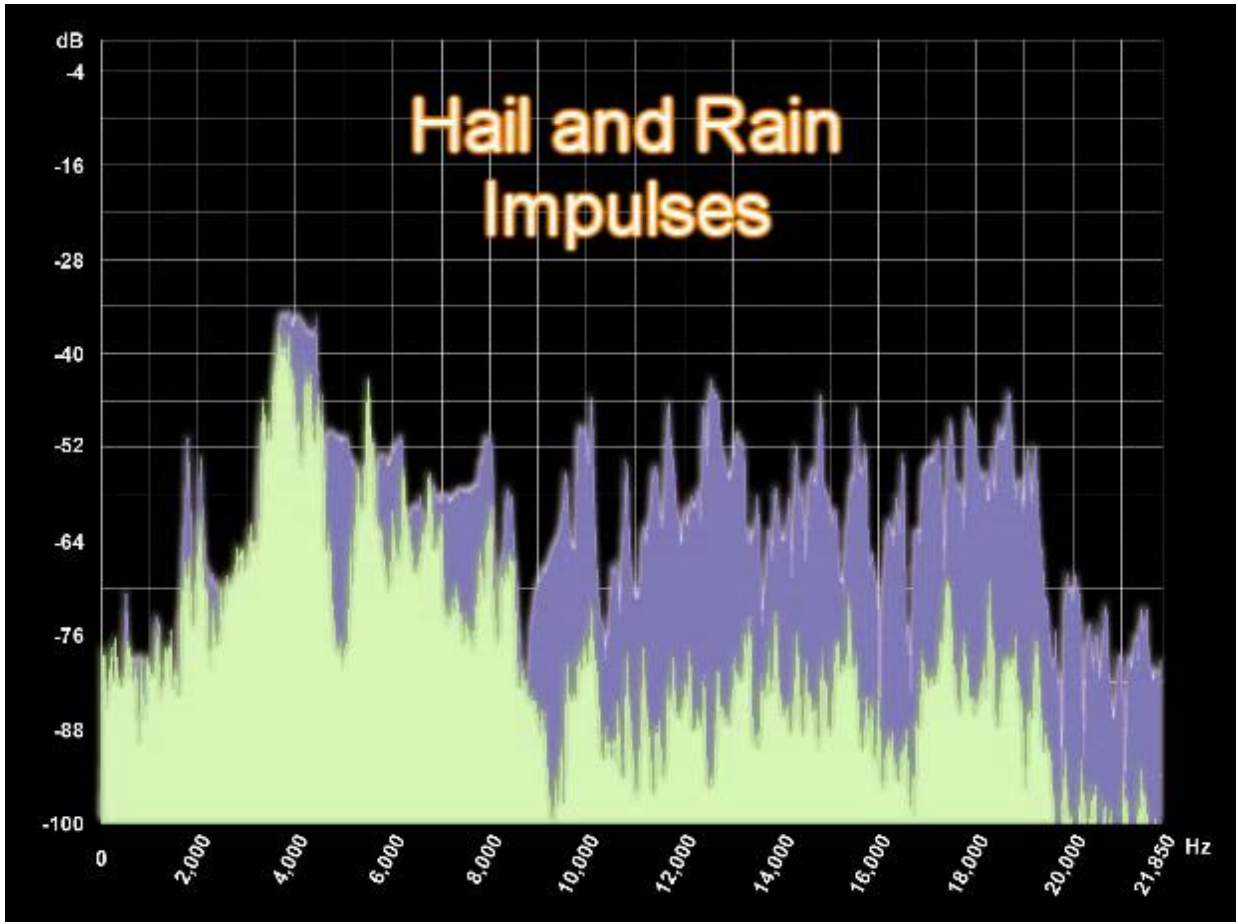

Impulse comparison.wmv

Movie 1. Frequency response of hail versus rain (video). The plot vertical axis is the magnitude (in dB) of the FFT of the hail transducer signal for single hail stone and rain drop hits. The horizontal axis represents frequency in Hz.

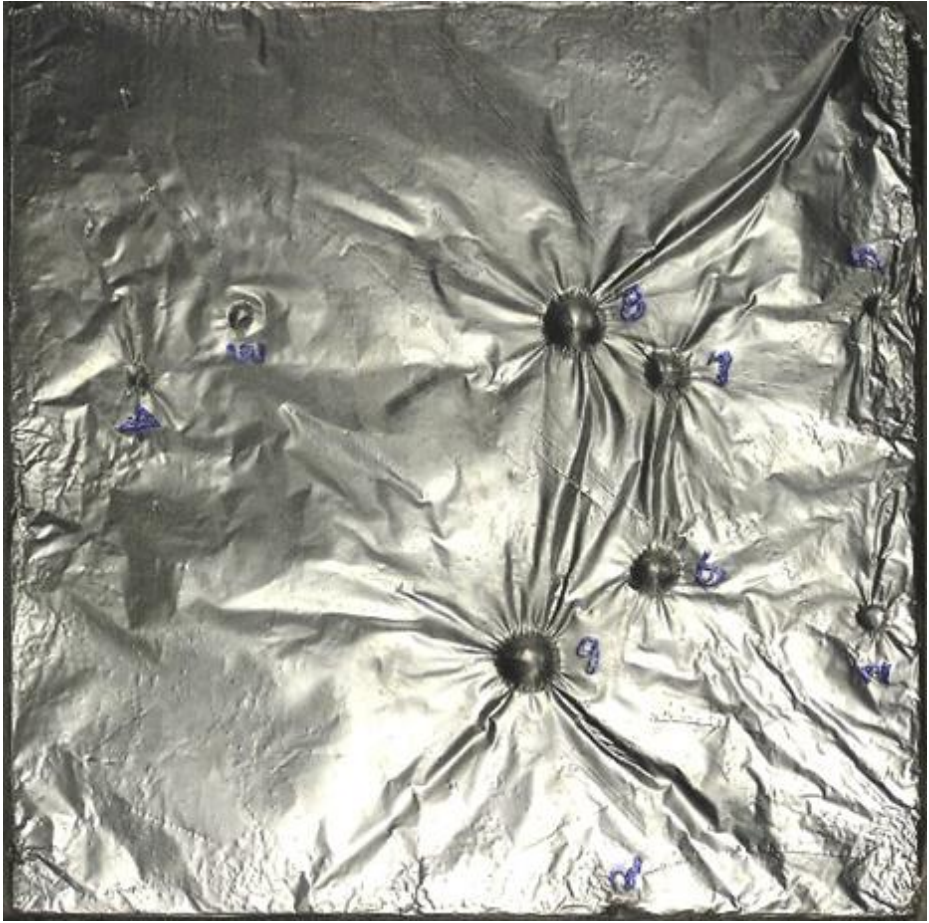

Figure 4. Hail pad dents from a 20-m drop test using various-size ice balls (numbers next to the dents mark the identity of the drop test).

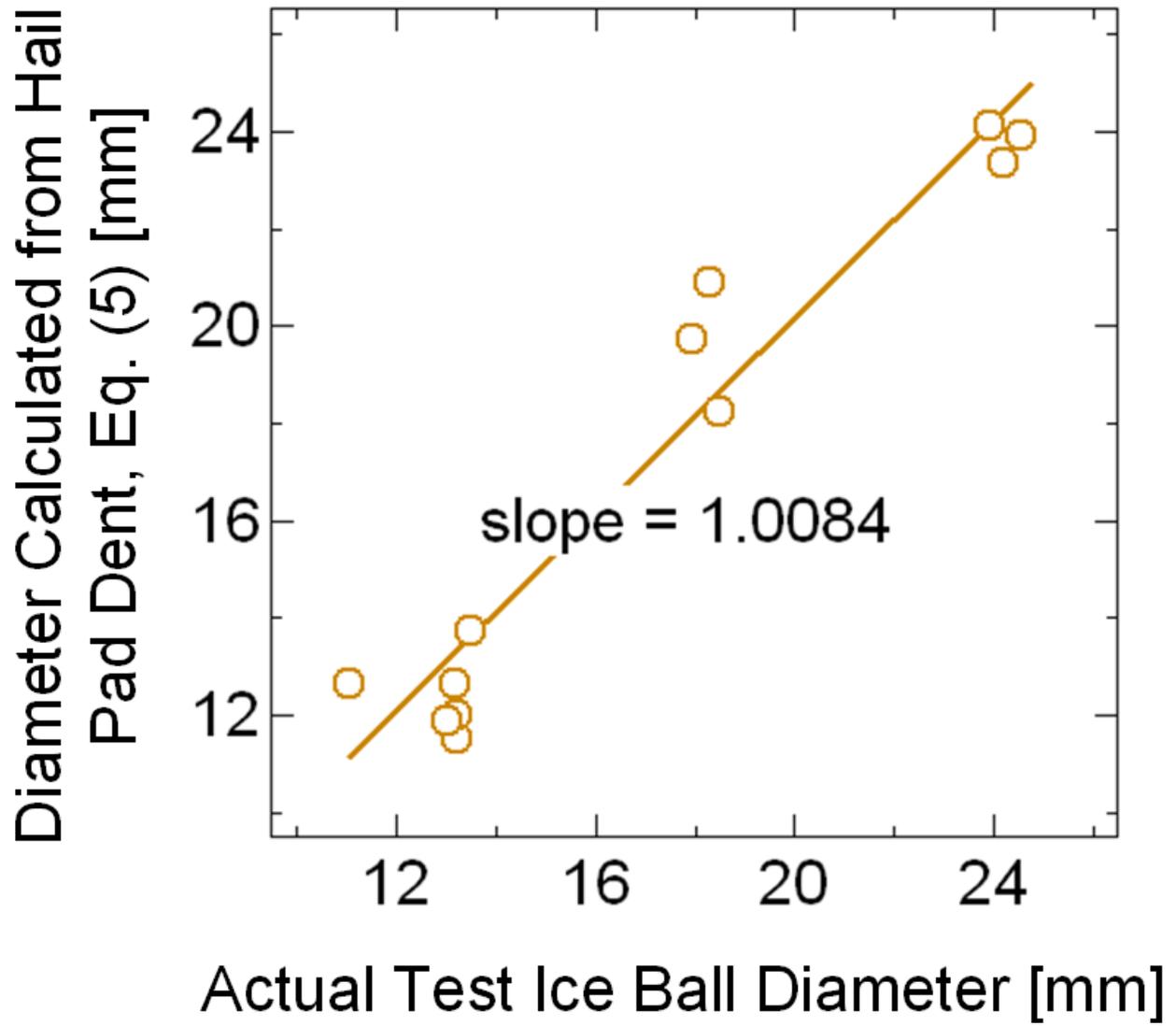

Figure 5. Hail pad calibration check using 20-m drop test data.

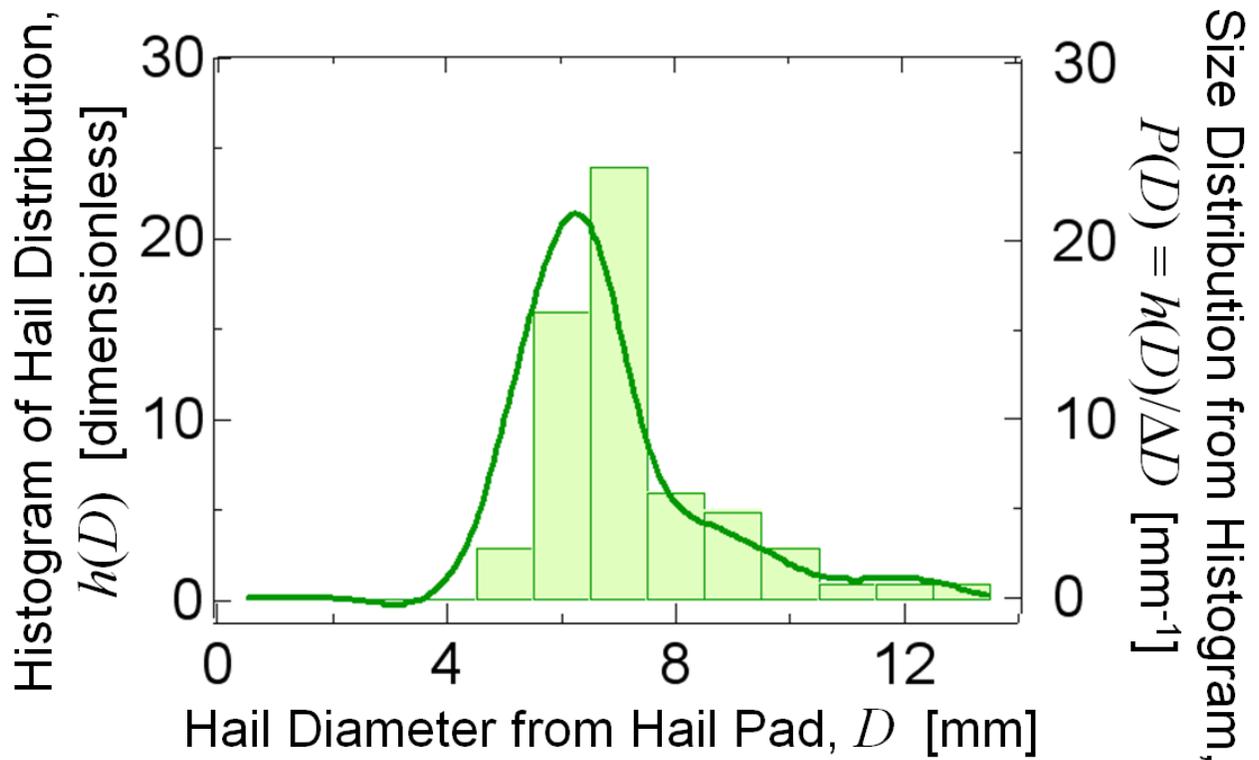

Figure 6. Hail diameters corresponding to hail pad for July 7, 2004, event. Solid line is PDF with $\Delta D = 1.0$ mm; bars are histogram.

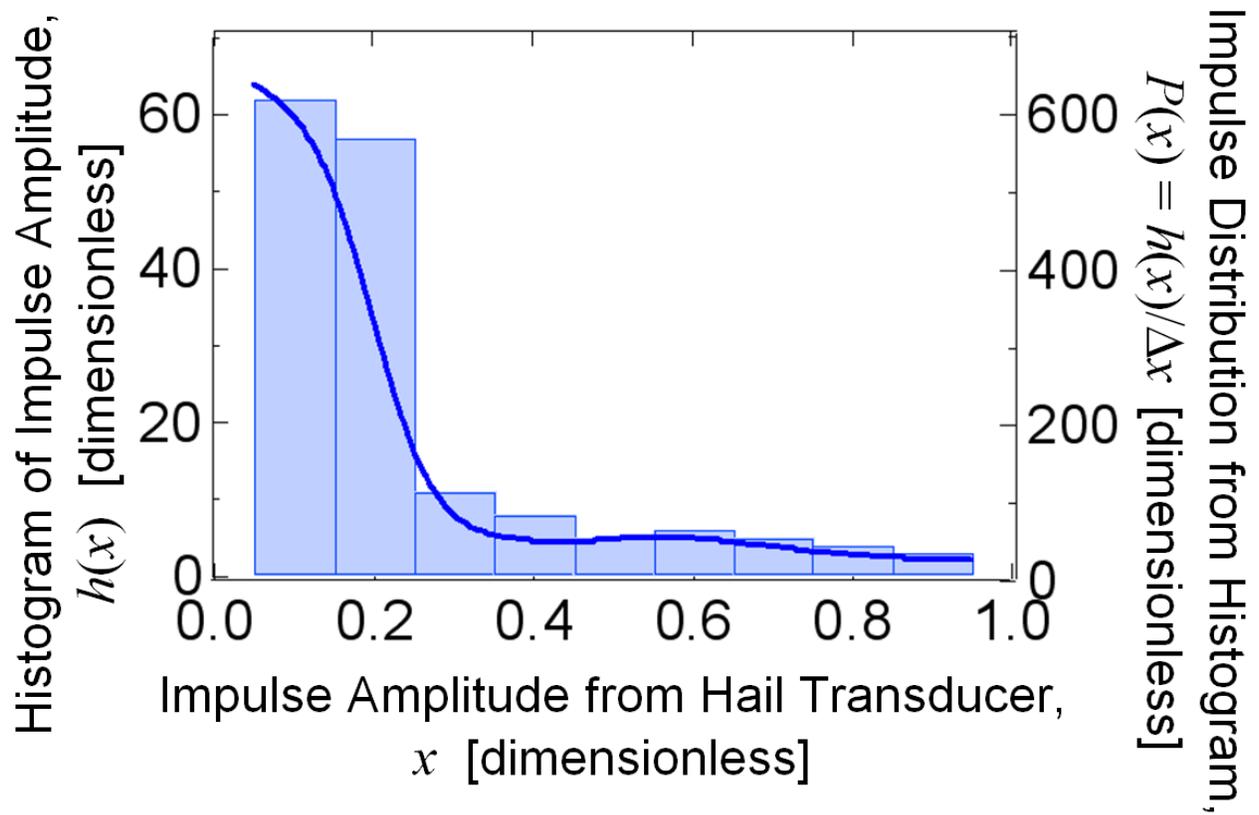

Figure 7. Hail transducer impulse amplitudes corresponding to July 7, 2004, event. Solid line is PDF with $\Delta x = 0.1$; bars are histogram.

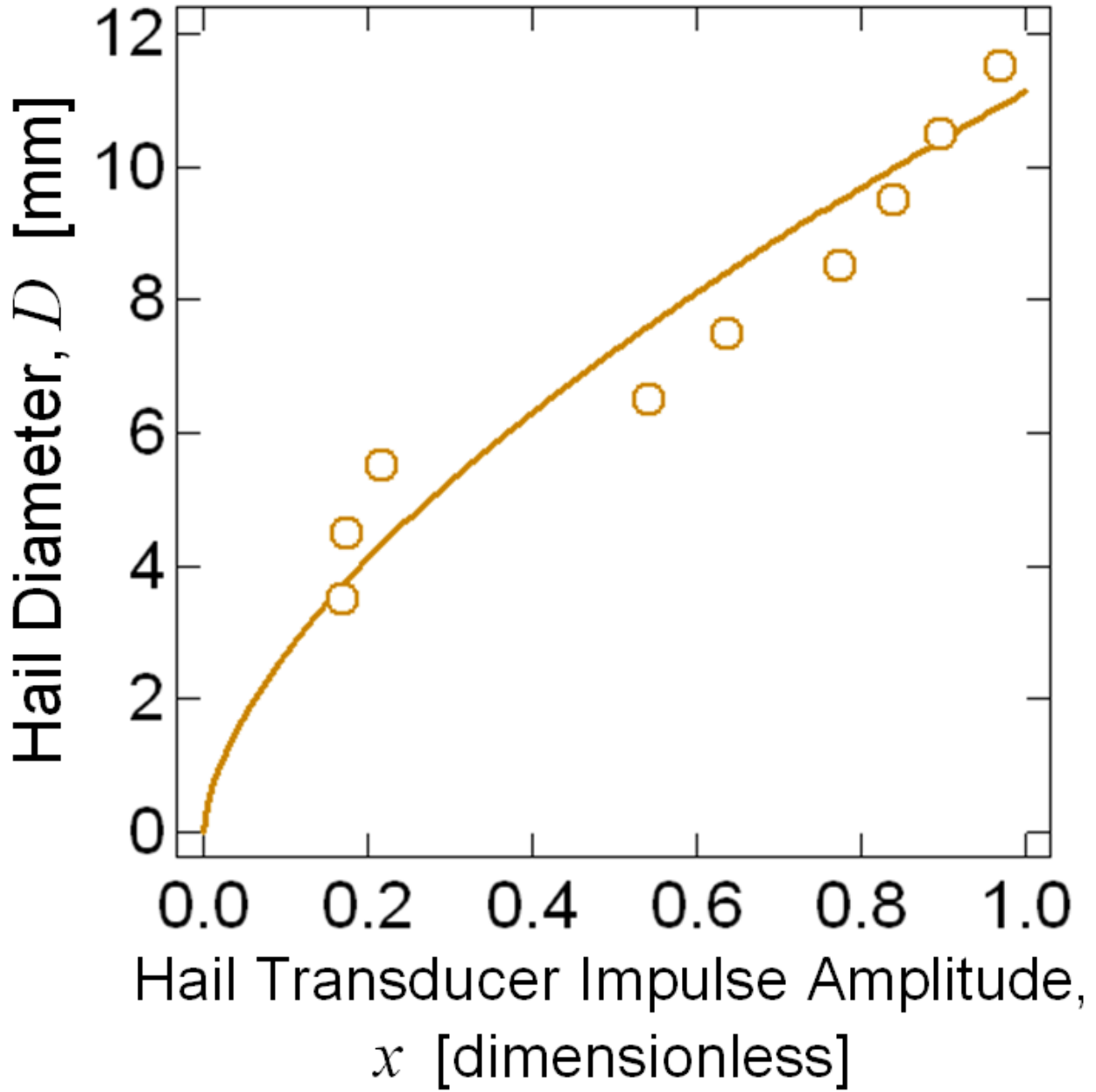

Figure 8. Hail transducer calibration curve based on the July 7, 2004 event. Solid line is fit from Eq. (8) with $A = 11.13$ and $b = 0.621$.

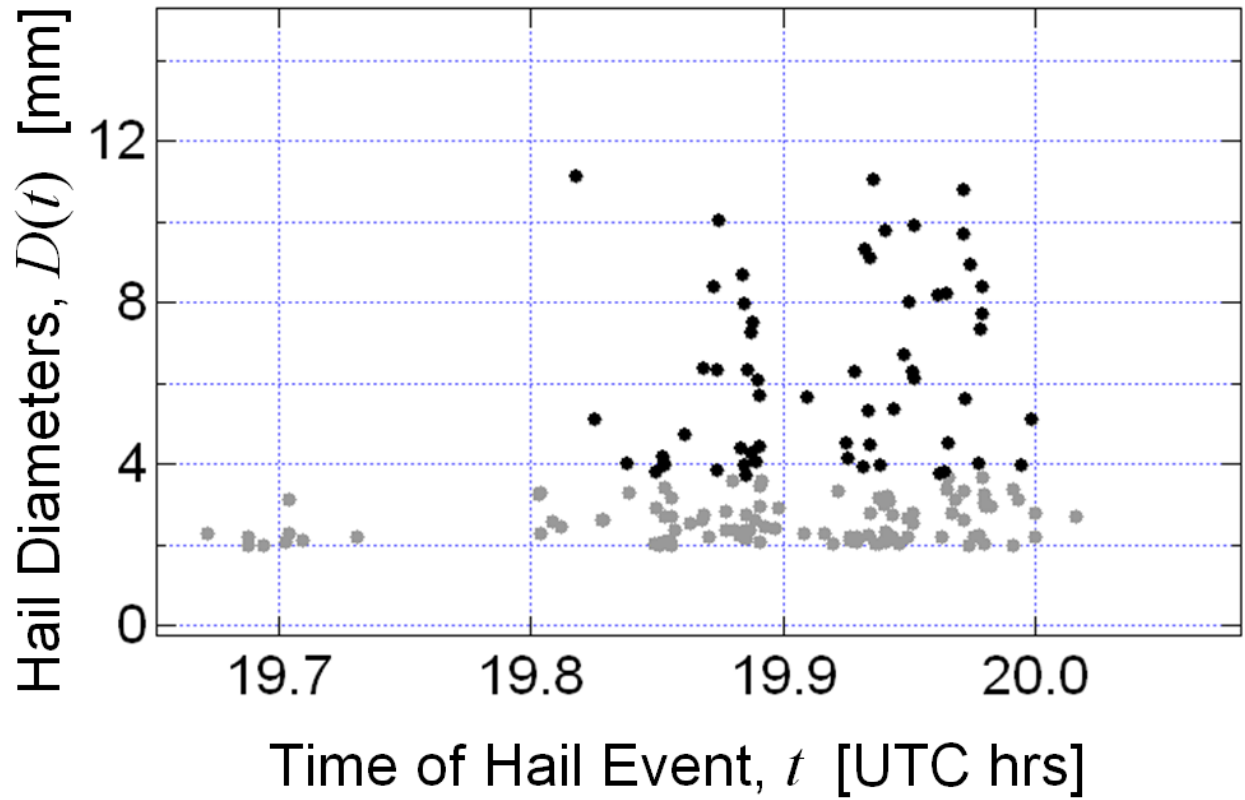

Figure 9. July 7 WEK hail data based on the calibration curve of Eq. (8). Gray circles correspond to extrapolated values below the hail pad $D_0$ threshold.